\documentclass[aps,prl,superscriptaddress,nobibnotes,twocolumn]{revtex4-1}
\usepackage[colorlinks=true,allcolors=black]{hyperref}
\usepackage[justification=raggedright,labelfont={bf},tableposition=top]{caption}
\usepackage{graphicx,braket,amsmath}
\setcitestyle{super}
\DeclareCaptionLabelSeparator{nature}{ $|$ }
\renewcommand{\figurename}{Figure}

\newcommand{\textmwone}[1]{\textcolor[rgb]{0.00,0.00,1.00}{#1}}
\newcommand{\textmwtwo}[1]{\textcolor[rgb]{0.00,1.00,1.00}{#1}}
\newcommand{\textrf}[1]{\textcolor[rgb]{1.00,0.00,0.00}{#1}}

\begin{document}

\title{Bell's inequality violation with spins in silicon}

\author{Juan P. Dehollain}
\thanks{These authors contributed equally to this work.}
\author{Stephanie Simmons}
\thanks{These authors contributed equally to this work.}
\author{Juha T. Muhonen}
\author{Rachpon Kalra}
\author{Arne Laucht}
\author{Fay Hudson}
\affiliation{Centre for Quantum Computation and Communication Technology}
\affiliation{School of Electrical Engineering and Telecommunications, UNSW Australia, Sydney NSW 2052, Australia}
\author{Kohei M. Itoh}
\affiliation{School of Fundamental Science and Technology, Keio University, 3-14-1 Hiyoshi, 223-8522, Japan}
\author{David N. Jamieson}
\author{Jeffrey C. McCallum}
\affiliation{Centre for Quantum Computation and Communication Technology}
\affiliation{School of Physics, University of Melbourne, Melbourne, VIC 3010, Australia}
\author{Andrew S. Dzurak}
\author{Andrea Morello}
\affiliation{Centre for Quantum Computation and Communication Technology}
\affiliation{School of Electrical Engineering and Telecommunications, UNSW Australia, Sydney NSW 2052, Australia}

%\date{\today}

%\begin{abstract}
%
%\end{abstract}

\maketitle

\textbf{Bell's theorem sets a boundary between the classical and quantum realms\cite{Bell1964p}, by providing a strict proof of the existence of entangled quantum states with no classical counterpart. An experimental violation of Bell's inequality demands simultaneously high fidelities in the preparation, manipulation and measurement of multipartite quantum entangled states. For this reason the Bell signal has been tagged as a single-number benchmark for the performance of quantum computing devices\cite{Rowe2001n,Ansmann2009n,Pfaff2012np}. Here we demonstrate deterministic, on-demand generation of two-qubit entangled states of the electron and the nuclear spin of a single phosphorus atom embedded in a silicon nanoelectronic device\cite{Muhonen2014nn}. By sequentially reading the electron and the nucleus, we show that these entangled states violate the Bell/CHSH\cite{Clauser1969prl} inequality with a Bell signal of 2.50(10). An even higher value of 2.70(9) is obtained by mapping the parity of the two-qubit state onto the nuclear spin, which allows for high-fidelity quantum non-demolition measurement (QND)\cite{Pla2013n} of the parity. Furthermore, we complement the Bell inequality entanglement witness with full two-qubit state tomography exploiting QND measurement, which reveals that our prepared states match the target maximally entangled Bell states with $>$96\% fidelity. These experiments demonstrate complete control of the two-qubit Hilbert space of a phosphorus atom, and show that this system is able to maintain its simultaneously high initialization, manipulation and measurement fidelities past the single-qubit regime.}

Bell's theorem provides a boundary to the strength of correlation that a pair of quantum two-level systems (qubits) can display, under the assumption that physical systems cannot be instantly affected by distant objects (``locality'') and that their properties exist before they are observed (``realism''). Certain quantum entangled states are predicted to violate Bell's theorem, and therefore invalidate local realistic interpretations of quantum mechanics\cite{Laloe2001ajp}. The most profound implications of Bell's theorem arise when observing entangled pairs of particles that are separated in space-time, such as photons travelling at the speed of light in different directions\cite{Aspect1982prl,Weihs1998prl}. Even for physical systems where space-time separation is not achievable, the Bell test carries an important message when placed in the context of using such systems for quantum information processing\cite{Bennett2000n}. It is in principle possible to produce and observe quantum entanglement without violating Bell's inequality, as is the case in a wide range of experiments, particularly in the solid state\cite{Ghosh2003n,Neumann2008s,Simmons2011n,Brunner2011prl,Shulman2012s,Bernien2013n}. However, a Bell's inequality violation constitutes an even stronger measure of the ability to faithfully produce, control and read out non-classical states of pairs of qubits\cite{Rowe2001n,Ansmann2009n,Pfaff2012np}. This maps directly onto the ability to perform high-fidelity entangling operations \cite{Peres1996prl}, which, along with single-qubit operations can fully access the full two-qubit Hilbert space.

Experimental access to Bell's theorem takes the form of the Clauser-Horne-Shimony-Holt (CHSH) inequality\cite{Clauser1969prl}. It involves the joint measurement of a two-qubit system along measurement axes $\alpha$ and $\beta$. The binary measurement outcomes ($0/1$) produce a correlation

\vspace{-12pt}
\begin{align}
E(\alpha,\beta) = P_{00} + P_{11} - P_{01} - P_{10} \label{eq:E}
\end{align}

where $P$ is the probability of detecting the subscripted measurement outcome. In its standard form, the inequality is tested by measuring each qubit along two axes, $\alpha, \alpha'$ and $\beta, \beta'$ respectively, and extracting $E$ in all four possible combination of axes. If the measurement setup does not allow for physically rotating the measurement axes, equivalent outcomes are obtained by rotating the qubit prior to a measurement along a fixed axis. The Bell signal is then

\vspace{-12pt}
\begin{align}
S = E(\alpha,\beta) + E(\alpha',\beta) + E(\alpha,\beta') - E(\alpha',\beta'). \label{eq:S}
\end{align}

Bell's theorem states that, within local realistic theories, $|S| \leq 2$. Conversely, quantum mechanics predicts $S_{max} = 2\sqrt{2}$ for a maximally entangled state and an appropriately chosen sets of axes.

\begin{figure*}[tb]
 \includegraphics[width=0.9\textwidth]{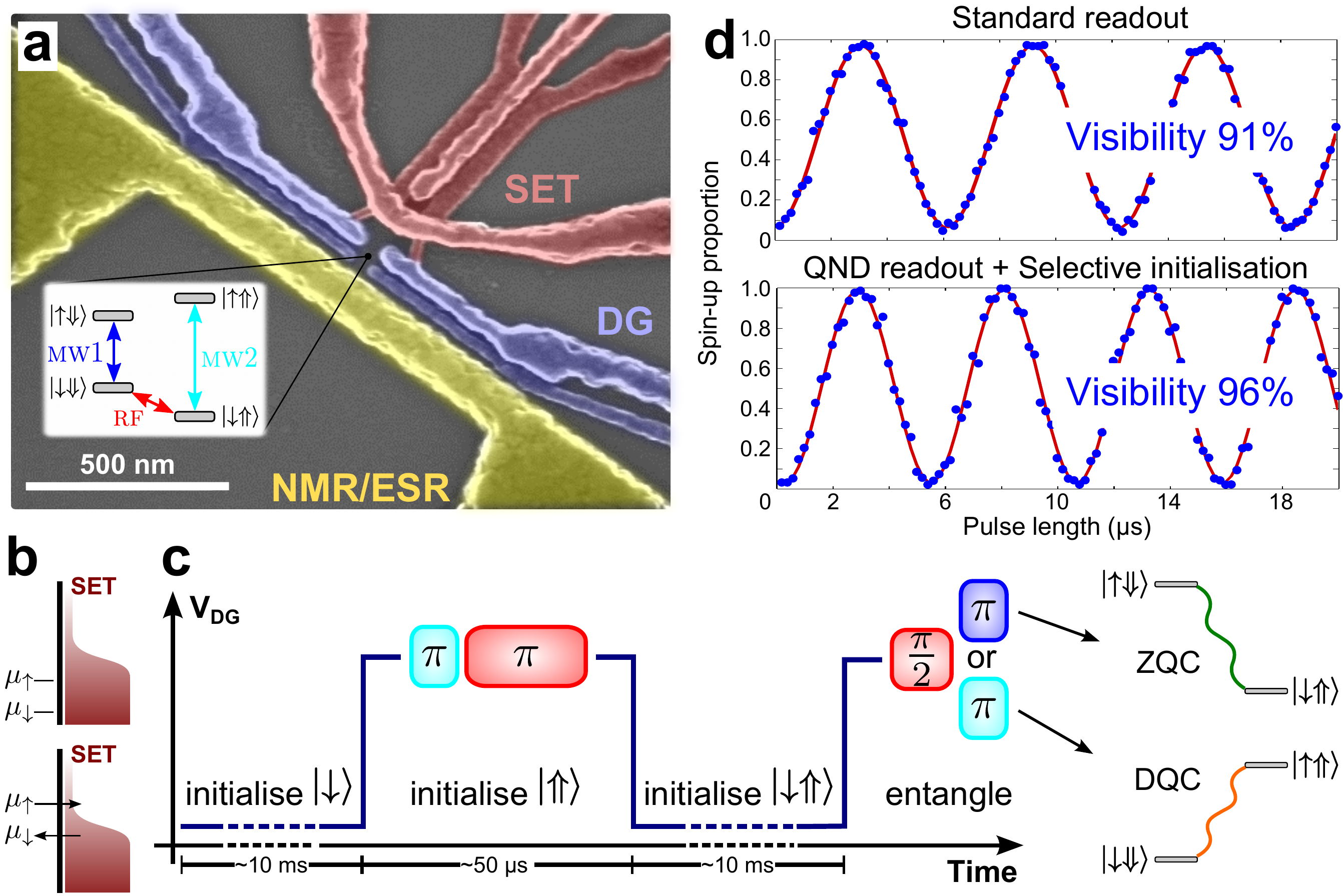}
 \caption{\textbf{Device operation and state preparation protocols.} \textbf{a.}~Coloured scanning electron microscope image of a device similar to the one used in the experiments. The gates coloured in red make up the single electron transistor (SET), used to perform sensitive charge sensing for electron spin readout; the donor gates (DG) coloured in blue, control the donor potential; the gate coloured in yellow is a broadband antenna used to drive the nuclear magnetic resonance and electron spin resonance (NMR/ESR) transitions. Inset shows a schematic four-level energy diagram of the two-spin electron-nuclear system under a static magnetic field. The colour coding of the three accessible transitions (${\textmwone{\textsc{mw}1}}$, ${\textmwtwo{\textsc{mw}2}}$, ${\textrf{\textsc{rf}}}$) will be used in the pulse sequences throughout the text and figures. \textbf{b.}(Bottom)~Electron $\ket{\downarrow}$ initialization is achieved by setting the donor potential such that $\mu_\uparrow>\mu_\textsc{set}>\mu_\downarrow$. Since $\gamma_eB_0 \gg k_{\rm B}T$, where $k_{\rm B}$ is the Boltzmann constant and $T \approx 100$~mK is the temperature of the electron reservoir, the ionized donor is predominantly neutralized by a $\ket{\downarrow}$ electron. (Top)~After initialization, $V_{\textsc{dg}}$ is increased to lower the donor potential to $\mu_{\downarrow}, \mu_{\uparrow} \ll \mu_{SET}$, preventing the electron from escaping the donor during the control phase. \textbf{c.}~The nuclear spin is initialized $\ket{\Uparrow}$ by applying the sequence $\pi_{\textmwtwo{\textsc{mw}2}}$:$\pi_{\textrf{\textsc{rf}}}$\cite{Dutt2007s}. If the nuclear spin is $\ket{\Uparrow}$, $\pi_{\textmwtwo{\textsc{mw}2}}$ flips the electron spin to $\ket{\uparrow}$, after which $\pi_{\textrf{\textsc{rf}}}$ is off-resonance and leaves the nuclear spin remains in the target state $\ket{\Uparrow}$. If the nuclear spin is $\ket{\Downarrow}$, $\pi_{\textmwtwo{\textsc{mw}2}}$ is off-resonance, allowing the subsequent $\pi_{\textrf{\textsc{rf}}}$ to flip the nuclear spin to $\ket{\Uparrow}$. A final electron $\ket{downarrow}$ initialization phase completes the $\ket{\downarrow\Uparrow}$ preparation. After the two-spin initialization sequence is performed, one of the maximally entangled states (DQC or ZQC) is prepared. \textbf{d.}~Electron Rabi oscillations highlighting the improvement in visibility with the implementation of selective initialization (see main text) and QND measurement protocols.\label{fig:intro}}
\end{figure*}

In the present experiment we use as qubits the electron ($\downarrow/\uparrow$) and the nuclear ($\Downarrow/\Uparrow$) spins of a single substitutional $^{31}$P donor, implanted\cite{Donkelaar2015jpcm} in an isotopically enriched $^{28}Si$ epilayer\cite{Muhonen2014nn,Itoh2014mrc}. The qubits are coupled by the hyperfine interaction $A\approx96.9$~MHz (shifted from the bulk value of 117~MHz due to the strong electric fields in the nanostructure\cite{Laucht2015sa}). A static magnetic field $B_0\approx1.55$~T induces a Zeeman splitting on the electron ($\gamma_eB_0$ with $\gamma_e\approx27.97$~GHz/T) and nuclear ($\gamma_\textsc{n}B_0$ with $\gamma_\textsc{n}\approx17.23$~MHz/T) spins, resulting in the two-qubit energy levels diagram depicted in the inset of \autoref{fig:intro}a. Manipulation of the full two-qubit Hilbert space requires access to at least three eigenstate transitions. Our experimental setup includes three microwave sources which deliver signals to a nanoscale broadband antenna\cite{Dehollain2013nt}, providing access to both electron spin resonance (ESR) transitions ($\nu_{\textsc{mw}\textmwone{1},\textmwtwo{2}} = \gamma_eB_0 \mp A/2$) and one nuclear magnetic resonance (NMR) transition ($\nu_{\textrf{\textsc{rf}}} = \gamma_\textsc{n}B_0 + A/2$). Coherent qubit rotations are denoted as $\alpha_s$ where $\alpha$ is the rotation angle in radians and the subscript identifies the transition. The $^{31}$P donor is located in the vicinity ($\approx 25$~nm) of the large electron island of a single-electron transistor (SET), formed by biasing a metal-oxide-semiconductor (MOS) gate stack above a SiO$_2$ layer. An additional set of gates (DG) controls the (spin-dependent) electrochemical potential of the donor, $\mu_{\uparrow,\downarrow}$, with respect to the SET potential $\mu_\textsc{set}$ (\autoref{fig:intro}a)\cite{Muhonen2014nn}. The SET island acts as a quasi-continuum electron reservoir for spin-dependent electron tunnelling to and from the donor, which is the key step for single-shot readout and electrical initialization of the electron qubit state\cite{Morello2010n}.

\begin{figure*}
 \includegraphics[width=0.9\textwidth]{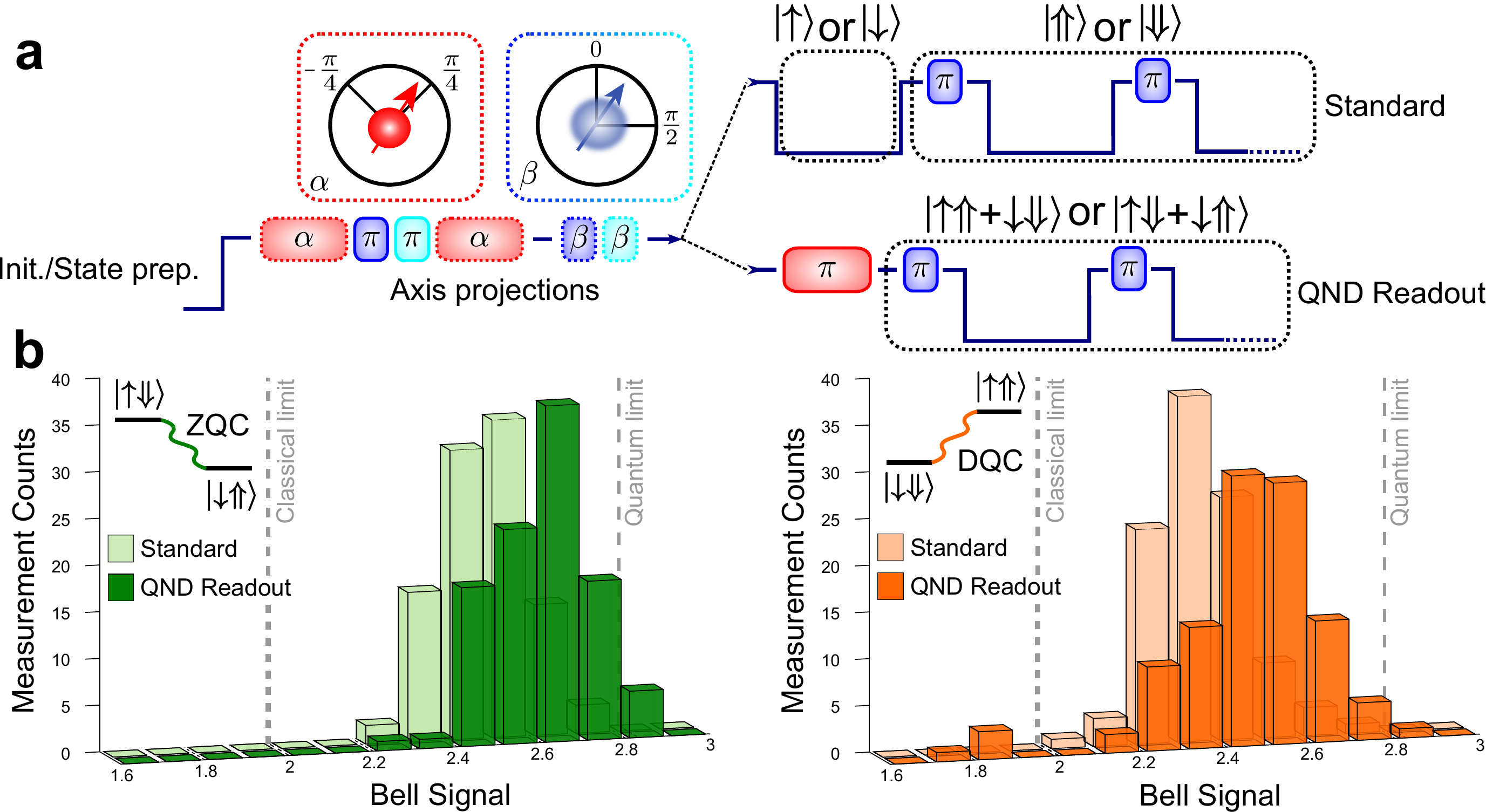}
 \caption{\textbf{Bell's inequality violation.} \textbf{a.}~Protocol for performing the CHSH experiment in the $^{31}$P electron-nucleus system. After preparation of an electron-nuclear entangled state, each qubit is rotated to obtain the desired combination of projection axes. Then, the correlation is extracted by measuring the qubits either directly or via mapping the parity onto the nuclear qubit, and performing a single QND readout on it. A correlation value $E(\alpha,\beta)$ is calculated after 300 repetitions of this sequence. \textbf{b.}~Histograms of Bell signal values obtained for the ZQC (left) and DQC (right) states. After obtaining $E$ for each of the 4 combinations of qubit rotations, a bell signal $S$ is calculated (\autoref{eq:S}). The histograms in the figure are constructed from $\sim100$~Bell signal values each. We assume a normal distribution to calculate the mean and standard deviation of $S$ displayed in \autoref{tab:bm}. \label{fig:bell}}
\end{figure*}

On-demand entanglement requires the two-spin system to be initialized to a known state. \autoref{fig:intro}b-c explains in detail the initialization procedure. The fidelity is limited by electron initialization errors. The probability of erroneous $\ket{\uparrow}$ electron initialization is proportional to the thermal population of electrons in the SET island at the potential $\mu_\uparrow$. Similarly, there is a small probability for a $\ket{\downarrow}$ electron to tunnel back to the SET, inversely proportional to the density of states in the SET island at $\mu_\downarrow$ \cite{Pla2012n}. Additionally, the finite ratio between the electron tunnel rate and the initialization time can be another source of error. If the donor is in the ionised state at the time the potential is lowered, the loading of the electron occurs while both spin states are equally accessible, resulting in a random initial electron spin state. To improve the initialization fidelity we monitor the SET current during the $\ket{\downarrow}$ initialization phase, and discard the traces where the donor is found ionised at the end of that phase (see Extended Data Figure \ref{edfig:init}). This selective initialization protocol improves the two-spin initialization fidelity from $94.3(7)\%$ to $\sim96.5(7)\%$.

We then apply the two-pulse sequence $\pi/2_{\textrf{\textsc{rf}}}$:$\pi_{\textmwone{\textsc{mw}1}}$ or $\pi/2_{\textrf{\textsc{rf}}}$:$\pi_{\textmwtwo{\textsc{mw}2}}$ to the $\ket{\downarrow\Uparrow}$ state to prepare a maximally entangled Bell state. By selecting ${\textmwone{\textsc{mw}1}}$ or ${\textmwtwo{\textsc{mw}2}}$ for the second pulse, we can prepare the zero quantum coherence (ZQC) $\ket{\Psi} = \frac{1}{\sqrt{2}}(\ket{\uparrow\Downarrow}+\ket{\downarrow\Uparrow})$ or the double quantum coherence (DQC) $\ket{\Phi} = \frac{1}{\sqrt{2}}(\ket{\uparrow\Uparrow}+\ket{\downarrow\Downarrow})$ states, respectively. The entire initialization and entangled state preparation sequence is depicted in \autoref{fig:intro}c.

\begin{figure*}
 \includegraphics[width=0.9\textwidth]{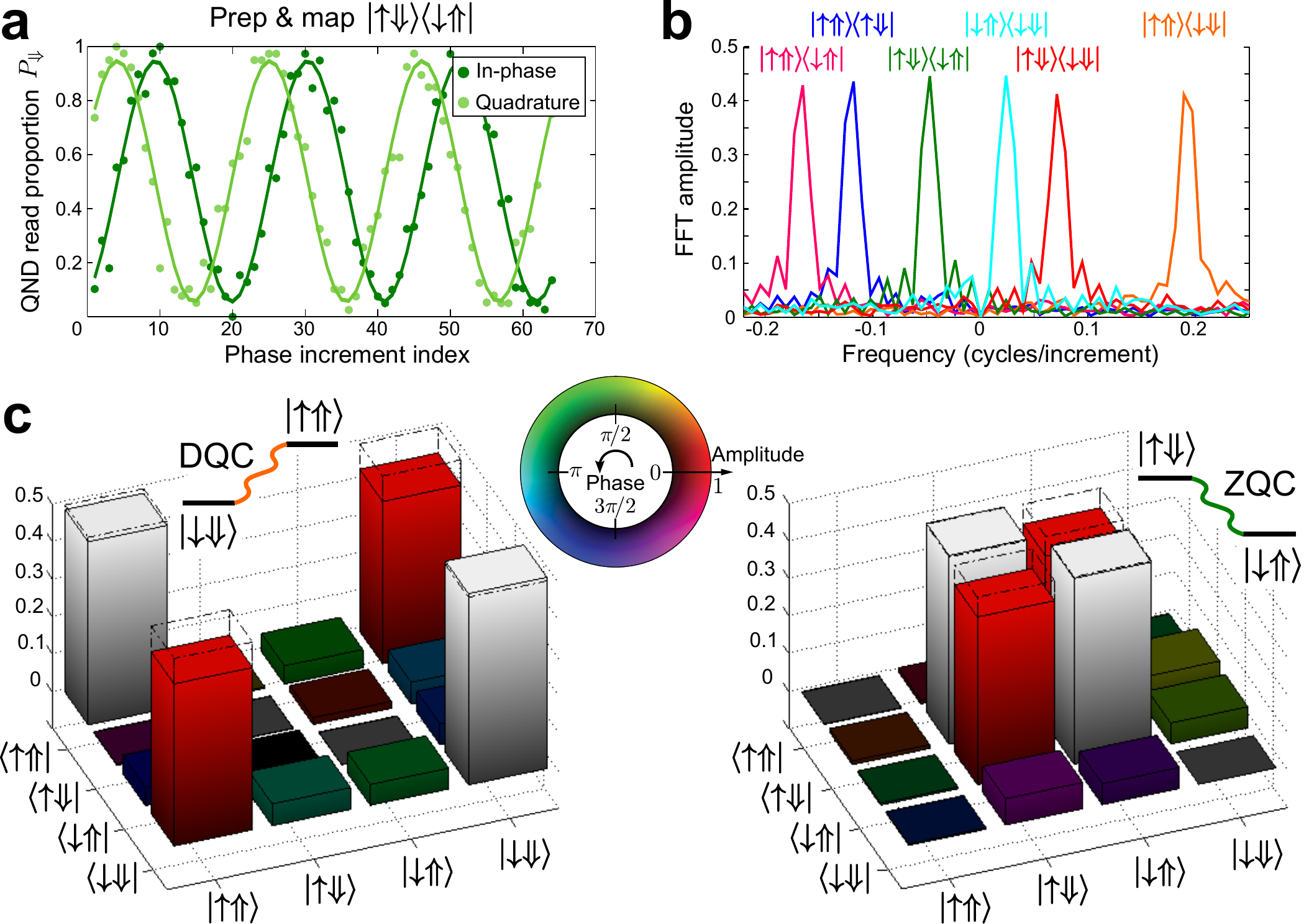}
 \caption{\textbf{Density matrix tomography.} \textbf{a.}~Example of state tomography through application of geometric phases. After preparing the ZQC state, we apply the geometric phase map, and then move the $\ket{\uparrow\Downarrow}\bra{\downarrow\Uparrow}$ coherence onto the nuclear spin observable, which is then measured in QND mode. The oscillations in the $\ket{\Downarrow}$ probability $P_{\Downarrow}$ arise from the increasing phase accumulated by the $\ket{\uparrow\Downarrow}\bra{\downarrow\Uparrow}$ coherence through the tomography protocol, and the visibility of the oscillations provides a lower bound on the amplitude of the $\ket{\uparrow\Downarrow}\bra{\downarrow\Uparrow}$ density matrix element. \textbf{b.}~Fourier transforms of tomography protocol signals obtained after preparing and mapping each of the off-diagonal coherences (protocols shown in Extended figure \ref{edfig:tomo1}). With carefully chosen phase increments ($\Delta\phi = -\frac{2}{21}\pi$ and $\Delta\theta = -\frac{5}{21}\pi$) we can obtain well separated peaks for each coherence. \textbf{c.}~Density matrices extracted for the DQC (Left) and ZQC (Right) states, expressed in polar coordinates where an off-diagonal element's phase is indicated by its colour. We have added a global phase correction to the matrices such that the maximum off-diagonal elements have zero phase.\label{fig:tomo}}
\end{figure*}

Characterization of the prepared entangled state requires high-fidelity single-shot projective measurements. The electron spin readout, based upon spin-dependent tunnelling from the donor to the SET island\cite{Morello2010n}, is destructive (the electron is lost after readout). Its fidelity, here up to $97\%$\cite{Muhonen2014nn}, is limited by the thermal broadening of the reservoir. In contrast, the nuclear qubit is measured through a quantum nondemolition (QND) method\cite{Pla2013n} by mapping its spin state onto that of the electron, through an electron-nuclear CNOT operation (e.g. by applying $\pi_{\textmwtwo{\textsc{mw}2}}$). The QND readout allows for repetitive measurement of the nuclear spin state, which reaches a fidelity $>99.9\%$ with 30 repetitions\cite{Muhonen2014nn}. With access to the full two-qubit Hilbert space, any observable in the four-level system can be mapped to the nuclear spin in order to take advantage of the high-fidelity QND measurement. To illustrate this point, we perform coherent Rabi rotations of the electron spin qubit. Their visibility is a nontrivial function of both measurement and initialization fidelities\cite{Pla2012n}. First we perform the measurement by direct electron readout, then by mapping the electron state onto the nucleus and performing QND readout on it. By further applying the initialization protocol described above, the overall Rabi visibility improves from $91\%$ to $96\%$ (\autoref{fig:intro}d), which means we have removed over half of the state preparation and measurement errors in the system.

The two-qubit system is now optimized to attempt CHSH inequality violations. The single qubit rotations required by the CHSH protocol are achieved through sequential rotations at each of the qubit's transitions. %This is a consequence of the fixed and relatively large two-qubit (hyperfine) interaction.
To perform a $\beta$ rotation on the electron independent of the nuclear state, we use the sequence $\beta_{\textmwone{\textsc{mw}1}}$:$\beta_{\textmwtwo{\textsc{mw}2}}$, since both electron frequencies (${\textmwone{\textsc{mw}1}}$ and ${\textmwtwo{\textsc{mw}2}}$) are simultaneously addressable. Conversely, for the nuclear spin we can only apply ${\textrf{\textsc{rf}}}$ within the $\ket{\downarrow}$ subspace. Therefore a $\alpha$ rotation on the nuclear qubit is performed through $\alpha_{\textrf{\textsc{rf}}}$:$\pi_{\textmwone{\textsc{mw}1}}$:$\pi_{\textmwtwo{\textsc{mw}2}}$:$\alpha_{\textrf{\textsc{rf}}}$, where the $\pi_{\textmwone{\textsc{mw}1}}$:$\pi_{\textmwtwo{\textsc{mw}2}}$ pulses bring the $\ket{\uparrow}$ manifold into the RF-addressable $\ket{\downarrow}$ subspace.

%In the qubit Bloch sphere picture,
A maximum Bell signal is achieved when the angle between the two projection axes on the same qubit (e.g. $\alpha$ and $\alpha'$) is $\pi/2$ and the relative shift between the two sets of axes is $\pi/4$. The correlation $E$ is obtained by independently measuring the individual qubits, through a single-shot readout of the electron spin immediately followed by a nuclear spin measurement. By repeating the sequence 300 times and logging each set of electron and nuclear spin results, we can construct the set of probabilities for each eigenstate $\{P_{\uparrow\Uparrow},P_{\downarrow\Downarrow},P_{\uparrow\Downarrow},P_{\downarrow\Uparrow}\}$ and compute $E$ from \autoref{eq:E}.

Alternatively, we can obtain $E$ in a single measurement by first mapping the parity of the Bell states (even parity for DQC, odd parity for ZQC) to the nuclear spin state, and then performing a QND measurement on it (See Extended Data Figure \ref{edfig:bell}a). This is done by applying a $\pi_{\textrf{\textsc{rf}}}$ pulse before a nuclear measurement, resulting in $\ket{\Uparrow}$ with probability $P_{\uparrow\Uparrow} + P_{\downarrow\Downarrow}$ and $\ket{\Downarrow}$ with probability $P_{\uparrow\Downarrow} + P_{\downarrow\Uparrow}$. Therefore, the correlation (Eq.~\ref{eq:E}) becomes $E = 1-2P_\Downarrow$. The complete sequence for the CHSH experiment---along with our chosen set of projected axes---is shown in \autoref{fig:bell}a.

The histograms in \autoref{fig:bell}b show clear violations of Bell's inequality for both DQC and ZQC states, using both standard and QND measurement protocols (see numbers in \autoref{tab:bm}). We obtain a maximum mean Bell signal of $2.70(9)$ for the ZQC state using QND readout, with the error taken as the standard deviation of the histogram data. Each histogram is constructed from a compilation of $\sim120,000$~single-shot measurements (see \autoref{fig:bell} and Extended Data Figure \ref{edfig:bell}b for the breakdown), taken over a period of $\sim11$~hours, highlighting the long-term stability of this system. The on-demand entangled state preparation and single-shot readout techniques adopted here also avoid the detection loophole\cite{Larsson2014jpa}.

In addition to the Bell test, we have experimentally mapped out the density matrix of the entangled states by adopting a tomography method first implemented by Mehring \textit{et al.}\cite{Mehring2003prl}, where the off-diagonal matrix elements of an electron-nuclear state are extracted from measurements of a single observable. We apply geometric phase operations to the input state and increment the geometric phase of the electron and nuclear spins relative to each other, through the sequence $\pi_{\textrf{\textsc{rf}}}^\phi$:$\pi_{\textmwtwo{\textsc{mw}2}}^\theta$, where $\phi$ and $\theta$ are the pulse rotation axes relative to the initial state preparation pulse. We sweep the phases by increments $\Delta\phi$, $\Delta\theta$ and map the desired off-diagonal coherence to the nuclear observable. The mapping is a pulse sequence that depends on the particular coherence being measured (see Extended Data Figure \ref{edfig:tomo1}). The measured $P_{\Uparrow}$ as a function of phase increment reveals oscillations with amplitude proportional to the off-diagonal density matrix element (\autoref{fig:tomo}a). The frequency of the oscillations is a linear function of $\Delta\phi$ and $\Delta\theta$, which can be chosen such that the recovered oscillations for each coherence have distinct frequencies (\autoref{fig:tomo}b).  The diagonal elements of the density matrix are obtained from the offsets of the tomography signals. Extended Data Figure \ref{edfig:tomo1} provides details on the pulse sequences, frequencies and offsets for each of the coherences of the system. The main advantages of this tomography method are twofold: first, the extracted coherence amplitudes are strict lower bounds and hence provide a conservative entanglement estimate; second, it only requires measurement of one of the qubits, which allows us to exploit the much higher fidelity of the nuclear qubit QND measurement. Additionally, this technique reduces false positive signal contributions which could arise from pulse errors. Each coherence evolves according to a distinct frequency, and when measuring a particular coherence only its particular frequency numerically contributes to the tomographic result; all other frequency contributions are due to pulse errors and are discarded (see Extended Data Figure \ref{edfig:tomo2}).

The density matrices (\autoref{fig:tomo}c, see Extended Data Figure \ref{edfig:tomo2} for numerical matrix) extracted using this protocol show remarkable fidelities $\mathcal{F}=\mathrm{Tr}\left[\sqrt{\sqrt{\rho_1}\rho_2\sqrt{\rho_1}}\right]>96\%$ when compared to the ideal states. Two additional density matrix tests, the positive partial transpose (PPT)\cite{Horodecki1996pla,Peres1996prl} and the concurrence ($\mathcal{C}$)\cite{Hill1997prl}, both confirm our measured states to be highly entangled. \autoref{tab:bm} shows all of the entanglement benchmarks obtained from both the Bell test measurements and density matrix tomography.

The deterministic preparation and detection of quantum entangled states that violate Bell's inequality with $S = 2.70(9)$ provides a striking demonstration of the potential of donor-based qubits in silicon for quantum information processing. In particular, it highlights the role of the nuclear spin in expanding the size of the available Hilbert space and in providing a pathway to high-fidelity QND measurements for all observables. Future experiments will further explore the role of nuclear spins in facilitating logic gates between electron qubits\cite{Kalra2014prx}, and in providing ancillas and long-lived memories in large fault-tolerant quantum computer architectures.

\begin{table}
\begin{tabular}{|c|*{2}{c}|*{3}{c}|}
\hline
& \multicolumn{2}{|c|}{\textbf{Bell Test}} & \multicolumn{3}{|c|}{\textbf{Tomography}} \\
State & Standard & QND & $\mathcal{F}$ & PPT & $\mathcal{C}$ \\
\hline
ZQC & 2.50(10) & 2.70(9) & 97(2)\% & -0.45(4) & 0.88(17) \\
\hline
DQC & 2.37(12) & 2.49(11) & 96(3)\% & -0.43(6) & 0.74(17) \\
\hline
\end{tabular}
\caption{Summary of electron-nuclear entanglement benchmarks.\label{tab:bm}}
\end{table}

\section*{Methods}
The fabrication process, cryogenic setup, gate biasing and data acquisition is identical to that described by Muhonen~\textit{et~al.}\cite{Muhonen2014nn}. We combine the signals of three different generators (Agilent E8257D/E8267D PSG for ESR and Agilent N5182 MXG for NMR) to apply pulses to each of the transitions described in the main text. We can achieve $\pi_{\textsc{mw}\textmwone{1},\textmwtwo{2}}$ rotations in $\sim3$~$\mu$s and $\pi_{\textrf{\textsc{rf}}}$ rotations in $\sim30$~$\mu$s. The generators are pulse modulated using a SpinCore PulseBlaster ESR TTL pulse generator. For phase control in the density matrix tomography experiments, we use the internal baseband arbitrary waveform generator (AWG) in the E8267D vector source, and we use an Agilent 81180A AWG to gate the I/Q inputs of the N5182.

%%% RERUN IF THERE ARE NEW REFERENCES %%%
%\bibliographystyle{natureJP}
%\bibliography{QC}
%\bibliography{QC,Papers}

\textbf{Acknowledgements.} This research was funded by the Australian Research Council Centre of Excellence for Quantum Computation and Communication Technology (project number CE110001027) and the US Army Research Office (W911NF-13-1-0024). We acknowledge support from the Australian National Fabrication Facility. The work at Keio was supported in part by the Grant-in-Aid for Scientific Research by MEXT, in part by NanoQuine, in part by FIRST, and in part by JSPS Core-to-Core Program.

\textbf{Author contributions.} J.P.D. and S.S. contributed equally to this work, designing and carrying out the experiments, and analysing the data, with A.M's supervision. J.P.D., J.T.M., A.L. and A.M. designed and constructed the experimental setup. R.K. and F.H. fabricated the device with A.S.D.'s supervision. K.M.I. supplied the $^{28}Si$ wafers. D.N.J. and J.C.M. designed and carried out the $^{31}P$ ion implantation with help from R.K.. J.P.D., S.S. and A.M. wrote the manuscript with feedback from all authors.

\textbf{Author information.} The authors declare no competing financial interests. Correspondence should be addressed to A.M.~(a.morello@unsw.edu.au).

\renewcommand{\figurename}{Extended Data Figure}
\setcounter{figure}{0}

\begin{figure*}
 \includegraphics[width=0.8\textwidth]{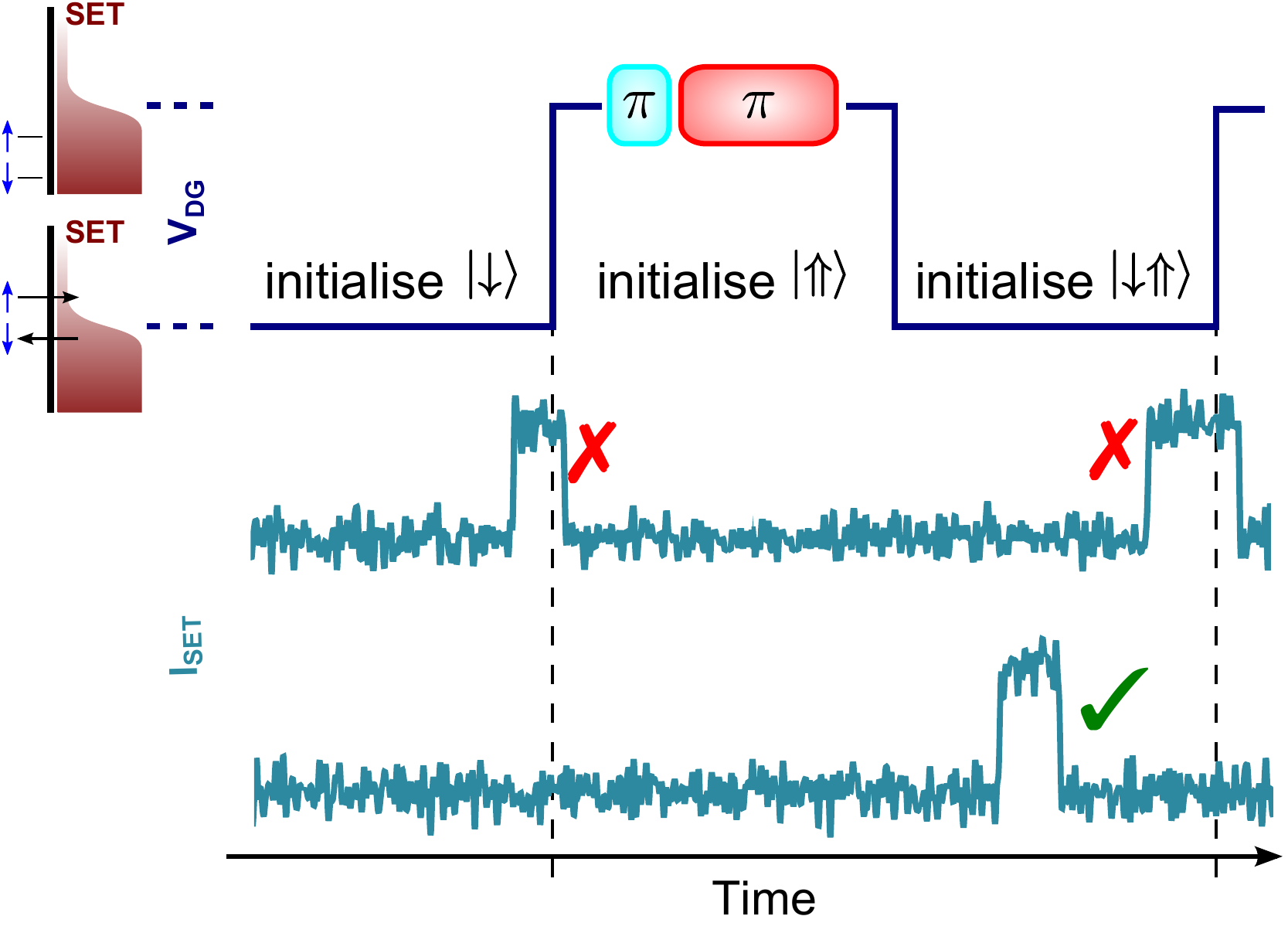}
 \caption{\textbf{Selective initialization protocol.} Top: pulsing scheme used to initialize the two-spin system in the $\ket{\downarrow\Uparrow}$ state, as described in the main text and Figure 1b-c. The SET current (blue traces) is tracked throughout the initialization sequence; high current implies ionized donor. The measurement is discarded if the donor is found to be ionized at the end of an electron $\ket{\downarrow}$ initialization phase.\label{edfig:init}}
\end{figure*}

\begin{figure*}
 \includegraphics[width=\textwidth]{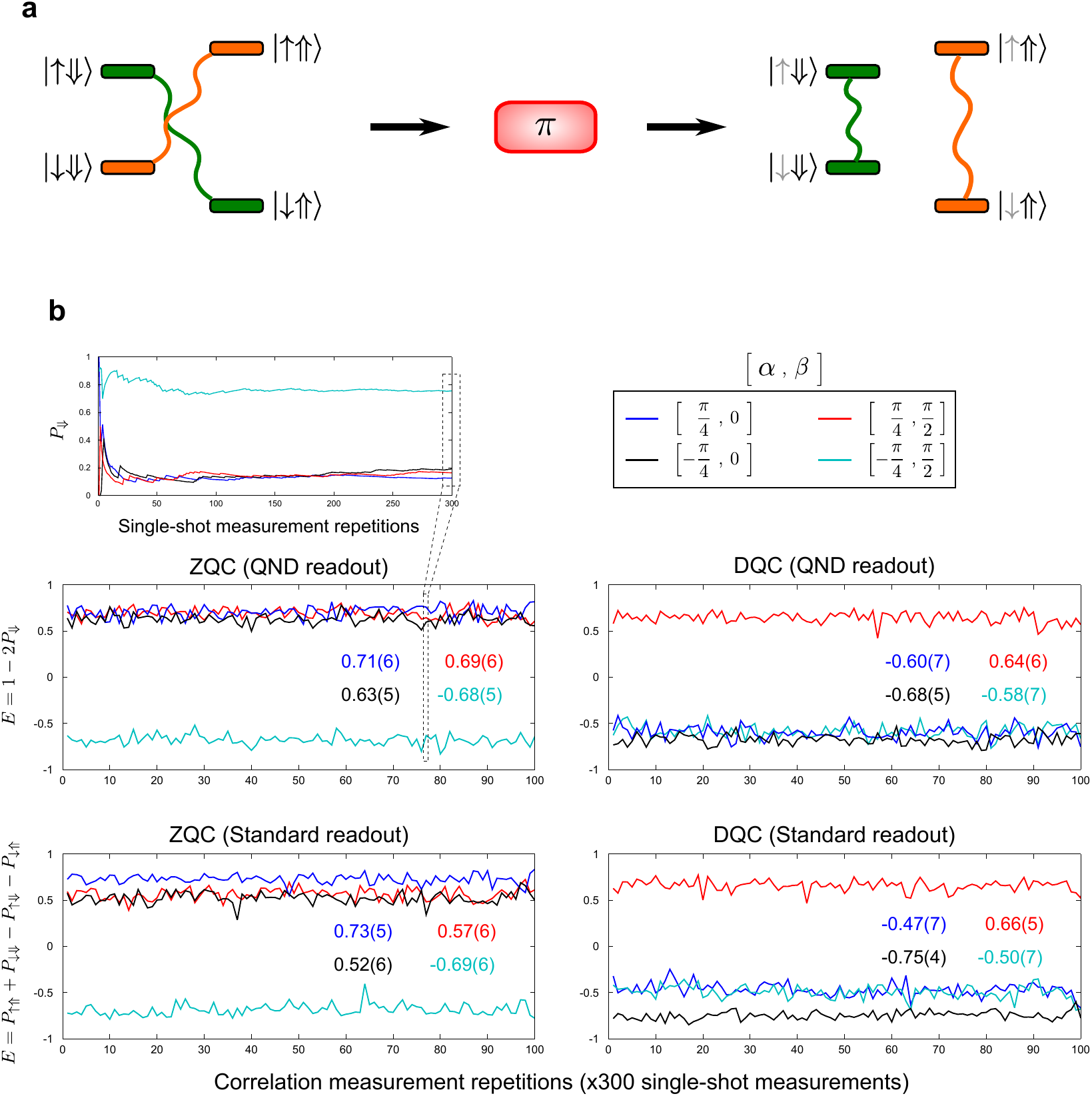}
 \caption{\textbf{Bell's inequality violation experiment details.} \textbf{a.} Diagram showing how a $\pi_{\textrf{\textsc{rf}}}$ maps the parity observable onto the nuclear observable. The pulse coherently swaps the coefficients of the $\ket{\downarrow\Downarrow}\leftrightarrow\ket{\downarrow\Uparrow}$ eigenstates, leaving the odd and even parity manifolds mapped on the $\ket{\Downarrow}$ and $\ket{\Uparrow}$ manifolds respectively. \textbf{b.} Correlation measurements used to construct the Bell signal histograms of Figure 2b. Line colours correspond to each of the nuclear and electron qubit projection combinations, as shown in the legend ($\alpha$ and $\beta$ are defined in Figure 2a). The top inset shows the cumulative moving average of $P_{\Downarrow}$ from a sample set of $300 \times 4$ single-shot measurements used to obtain one correlation set. For the QND measurements, we calculate the correlation $E = 1-2P_\Downarrow$ after each set of 300 single-shot measurements.\label{edfig:bell}}
\end{figure*}

\begin{figure*}
 \includegraphics[width=\textwidth]{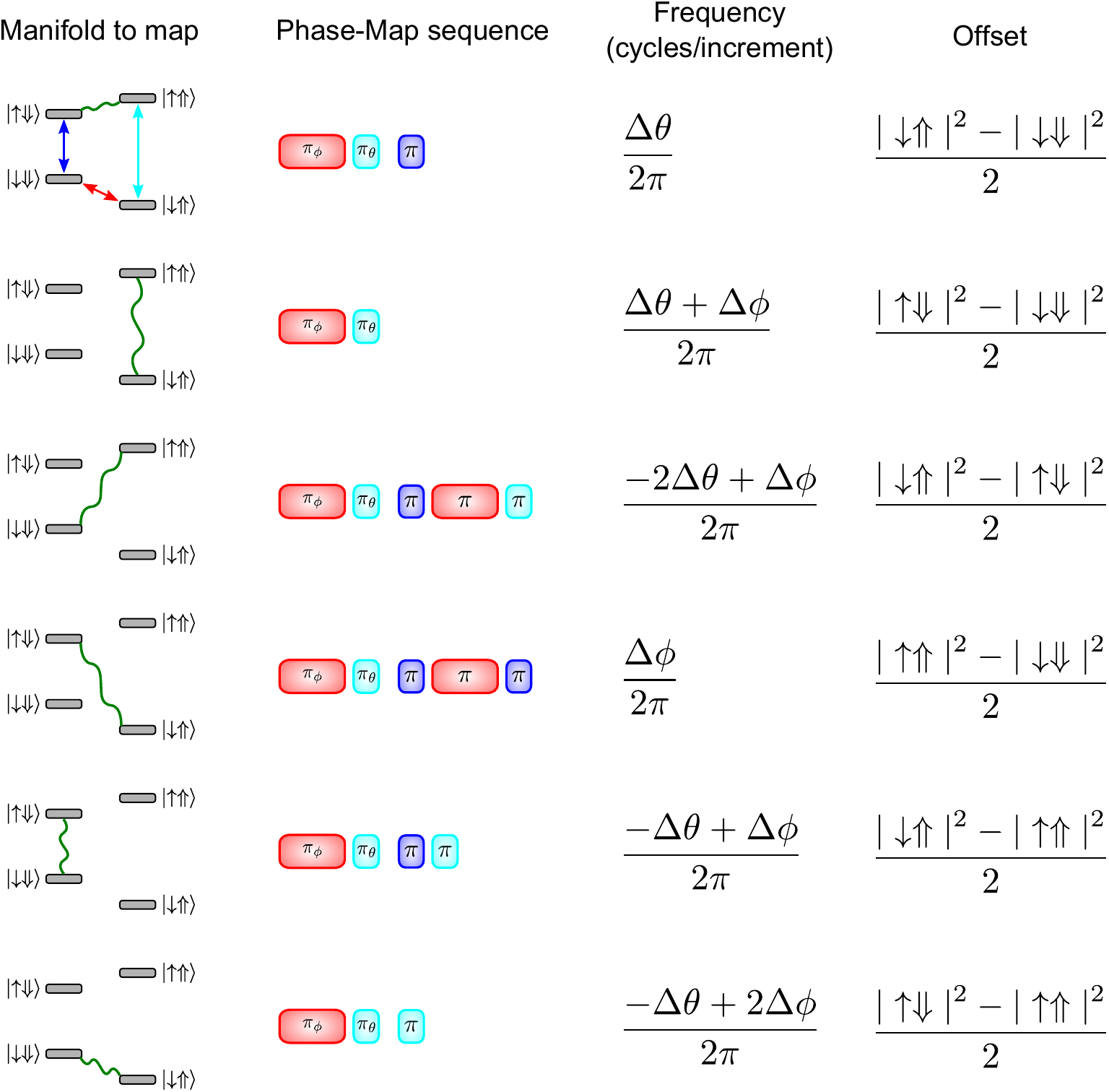}
 \caption{\textbf{Density matrix tomography pulse sequences and fitting parameters.} Left column: set of coherences measured by the tomography protocol. After preparing each tomography target state, a $\pi_{\textrf{\textsc{rf}}}$ and $\pi_{\textmwtwo{\textsc{mw}2}}$ pulse are applied with phase offsets $\phi$ and $\theta$ respectively. These pulses act as phase gates, which alter the phase of the coherence. The rest of the pulse sequence maps the coherence to be measured onto the $\ket{\downarrow\Downarrow}\bra{\downarrow\Uparrow}$ manifold (column 2). A $\pi/2_{\textrf{\textsc{rf}}}$ pulse is then applied to project the coherence onto the nuclear observable before performing a QND nuclear measurement. The nuclear spin proportion is plotted as we increment the phase offsets by $\Delta\phi$ and $\Delta\theta$, and the resulting signals (see Extended Data Figure 4) are fitted to $P_\Downarrow(n_{ph}) = A\sin(2\pi f_p n_{ph} + B) + C + 0.5$. Here, $n_{ph}$ is the phase increment number and $f_p$ is a frequency in cycles/increment which is unique to each coherence (column 3). $A$ and $B$ are free fitting parameters from which we extract the off-diagonal element amplitude and phase respectively. The remaining free fitting parameter $C$ is the offset of the measured signal, which gives information on the eigenstate populations (column 4). The diagonal elements of the density matrix are extracted by constructing a system of equations with all of the extracted offsets and $|\uparrow\Uparrow|^2+|\uparrow\Downarrow|^2+|\downarrow\Uparrow|^2+|\downarrow\Downarrow|^2=1$. The overdetermined system is solved using a non-negative least-squares solving algorithm.\label{edfig:tomo1}}
\end{figure*}

\begin{figure*}
 \includegraphics[width=0.9\textwidth]{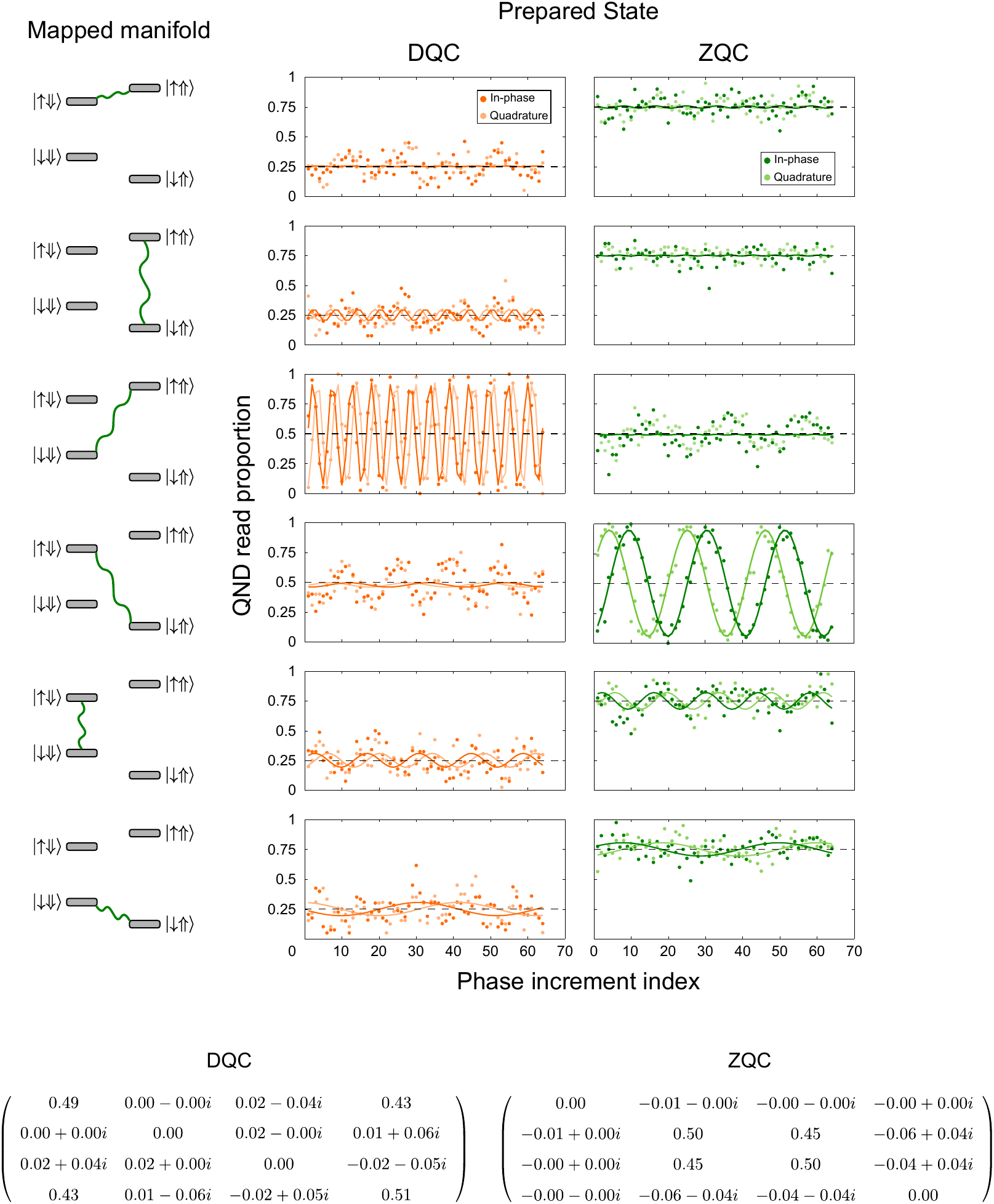}
 \caption{\textbf{Density matrix tomography detailed results.} Plots correspond to the resulting signals obtained from following the density matrix tomography protocol for each of the coherences, after preparing the entangled states DQC and ZQC. The data (dots) is fitted (solid line) as described in Extended Data Figure 3. The quadrature component is obtained by applying a $\pi/2$ phase offset to the final projective $\pi/2_{\textrf{\textsc{rf}}}$ pulse and is fitted by adding $\pi/2$ to the $\sin$ argument in the fitting function. Note that some traces display oscillations at frequencies which are different from the characteristic frequency of the coherence. These oscillations arise from cross-talk between coherences that can result from the tomography pulse sequence. The cross-talk is minimized by carefully choosing the rate of phase increments for different coherences, to ensure that their Fourier transforms are well spaced from each other. Black dashed lines indicate offsets for the ideal states. The resulting density matrices for each entangled state are presented in numerical form at the bottom. We assume the matrix is Hermitian and apply a global phase correction so the coherence with greatest amplitude has zero phase.\label{edfig:tomo2}}
\end{figure*}

\end{document}